\documentclass[preprint]{elsarticle}
\pdfoutput=1 

\usepackage{subfigure}
\usepackage{xcolor}
\usepackage[T1]{fontenc} 
\usepackage{xspace}
\usepackage{multirow}
\usepackage{hyperref}
\usepackage{amsfonts}
\usepackage{amssymb}
\usepackage{amsmath}

\usepackage[utf8]{inputenc}

\DeclareMathOperator{\sech}{sech}

\def\GeV{\ifmmode {\mathrm{\ Ge\kern -0.1em V}}\else \textrm{Ge\kern -0.1em V}\fi}%
\def\GeV{\ifmmode {\mathrm{\ Ge\kern -0.1em V}}\else \textrm{Ge\kern -0.1em V}\fi}%

\title{\boldmath Optical effects of domain walls}

\author[1]{Valentin V. Khoze}
\ead{valya.khoze@durham.ac.uk}
\author[1]{Daniel L. Milne}
\ead{daniel.l.milne@durham.ac.uk}

\affiliation[1]{organization={IPPP}, 
                            addressline={Department of Physics, Durham University},
                            city={Durham}, 
                            postcode={DH1 3LE},
                            country={UK}}

\begin{document}

\begin{abstract}
Domain walls arise in theories where there is spontaneous symmetry breaking of a discrete symmetry such as $\mathbb{Z}_{N}$ and are a feature of many BSM models. In this work we consider the possibility of detecting domain walls through their optical effects and specify three different methods of coupling domain walls to the photon. We consider the effects of these couplings in the context of gravitational wave detectors, such as LIGO, and examine the sensitivity of these experiments to domain wall effects. 
In cases where gravitational wave detectors are not sensitive we examine our results in the context of axion experiments and show how effects of passing domain walls can be detected at interferometers searching for an axion signal.
\end{abstract}

\maketitle


\section{\label{Sec:Intro}Introduction}

It was pointed out in Ref.~\cite{Jaeckel:2016jlh} that gravitational wave detectors, such as LIGO, can be used to search for potential signals of cosmological domain walls~\cite{Sikivie:1982qv}. The LIGO detector was built with the purpose of testing general relativity by searching for gravitational waves \cite{Abbott:2016blz}. However, it was shown by Jaeckel et al. that domain walls could leave an optical signal at an interferometer-based experiment that is not caused by gravity but is triggered instead by an effective interaction between the domain wall and the photon.
Domain walls arise naturally in theories when a discrete symmetry is broken; such discrete symmetries occur frequently in many beyond the Standard Model (BSM) settings in particle phyisics and are readily seen in the literature. Simple $\mathbb{Z}_{2}$ symmetries are popular in the context of two Higgs doublet models \cite{Branco:2011iw, Belanger:2013xza} and aspects of domain walls in these models were specifically studied in \cite{Battye:2020jeu,Battye:2020sxy}. Many other discrete symmetries arise in many different areas of physics such as flavour physics \cite{Altarelli:2010gt} and dark matter \cite{Baer:2014eja}. 

\medskip

In this paper we extend on the work of \cite{Jaeckel:2016jlh} by considering different types of interactions between the domain walls sector and QED and also accounting for the effect of cavities in interferometers. We couple the domain wall to the photon in three different ways and show that in each case the signal in an interferometer is quite different and, in fact, an experiment sensitive to one type of coupling is not in general sensitive to the others. It should be emphasised that many previous papers have considered gravitational waves in the context of domain walls, see e.g. Refs.~\cite{Nakayama:2016gxi,Jaeckel:2020mqa,Saikawa:2017hiv}, but in general these earlier studies considered gravitational waves sourced by annihilations between domain walls. In this paper we do not examine gravitational waves themselves but the signal at a gravitational wave interferometer which is actually generated by the interaction of the domain wall and the photon and not by the distortion of the Minkowski metric.

\begin{figure*}
\includegraphics[width=0.5\textwidth]{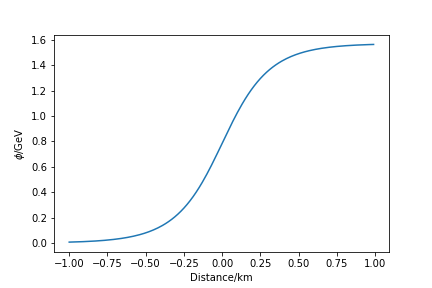}
\includegraphics[width=0.5\textwidth]{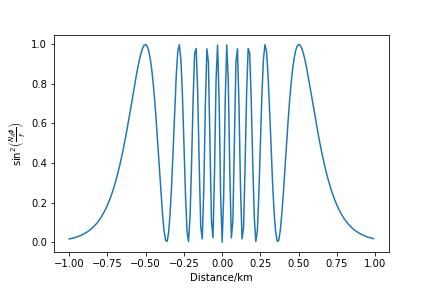}
\caption{The profile of the domain wall field \eqref {eq:DWsolution} (left) and the function $\sin^{2}\left(\frac{N_{A}\phi}{f}\right)$ (right) plotted as functions of the distance $z$ for $m=1$ neV, $f=1$ GeV
 and $N_{A}=20,\,N_{\phi}=4$.}
\label{fig:DW}
\end{figure*}

For the domain wall itself we will use the generic simple model of Ref.~\cite{Pospelov:2012mt},  
\begin{equation}
\label{eq:Lagrangian}
\mathcal{L}_{\rm DW}=\frac{1}{2}\left(\partial_{\mu}\phi\right)^{2}-\frac{2m^{2}f^{2}}{N_{\phi}^{2}}\sin^{2}\left(\frac{N_{\phi}\phi}{2f}\right)
\end{equation}
which has domain wall solutions,
\begin{equation}
\label{eq:DWsolution}
\phi_{\rm cl}\left(z\right)=\frac{4f}{N_{\phi}}\tan^{-1}\left(e^{mz}\right).
\end{equation}
The Lagrangian \eqref{eq:Lagrangian} is invariant under the shift symmetry $\phi \to \phi + \frac{2 \pi}{N_{\phi}} \,f$ which is the consequence of the 
$\mathbb{Z}_{N_{\phi}}$ symmetry of the underlying theory of a complex scalar field $\Phi (x) = S(x)\,e^{i \phi(x)/f}$.
Since the domain wall field $\phi$ is the phase of a complex scalar $\Phi$, it is a pseudo-scalar, and hence the model \eqref{eq:Lagrangian} and the corresponding classical solution \eqref{eq:DWsolution} describe pseudo-scalar (and for small $m$) light domain walls.
The plot on the left in Fig.~\ref{fig:DW} shows the domain wall solution \eqref{eq:Lagrangian} as a function of the normal distance $z$ to the domain wall in km.

\medskip

We will consider three types of interactions between the domain wall field and the photon field in QED that encompass couplings to all photon operators of dimension $\le 4$. These correspond to
 an effective photon mass-term coupling, $\mathcal{L}\supset \frac{1}{2} m_0^2\,\sin^{2}\left(\frac{N_{A}\phi}{f}\right)A_{\mu}A^{\mu}$, 
 an axion-like coupling, $\mathcal{L}\supset \frac{1}{4} \tilde{g}_{DW}\,(\phi/f) \, F_{\mu\nu}\tilde{F}^{\mu\nu}$ and a coupling to the canonical photon kinetic term, $\mathcal{L}\supset \frac{1}{4}  g_{DW}\, (\phi /f)^2\, F_{\mu\nu}F^{\mu\nu}$. The photon mass-term coupling is considered in Section 2, the domain wall axion-like coupling is dealt with in Section 3, and in Section 4 we compute the effects of the photon kinetic term coupling to domain walls. Conclusions are presented in Section 5.

\section{\label{Sec:massterm}Signal from a photon mass term coupling}

In this section, we review the results of \cite{Jaeckel:2016jlh} and extend this work by accounting for the Fabry-Perot cavities used in LIGO. The interaction of the domain wall field $\phi$ with the lowest dimension ($d=2$) photon field operator is given by the effective mass term for the photon sourced by the domain wall field. We thus consider the modified QED Lagrangian,
\begin{equation}
\mathcal{L}=\mathcal{L}_{QED}+\frac{1}{2}m_{0}^{2}\sin^{2}\left(\frac{N_{A}\phi}{f}\right)A^{\mu}A_{\mu} \,.
\end{equation}
The mass parameter $m_0$ will ultimately be taken to be in the sub-neV regime and the 
domain wall-induced function $\sin^{2}\left(\frac{N_{A}\phi}{f}\right)$ is plotted in Fig.~\ref{fig:DW} (plot on the right).
Note that computed on the domain wall solution, this function vanishes far from the centre of the domain wall as long as $N_{A}/N_{\phi}$ is integer or half-integer, so that the photon mass is non-vanishing only in the vicinity of the domain wall. This changes the dispersion relation of light which becomes,
\begin{equation}
\label{eq:massdisrel}
\omega^{2}=k^{2}+m_{0}^{2}\sin^{2}\left(\frac{N_{A}\phi}{f}\right),
\end{equation}
where $k=|{\bf k}|$ is the magnitude of the photon 3-momentum.

\begin{figure}
\includegraphics[width=\columnwidth]{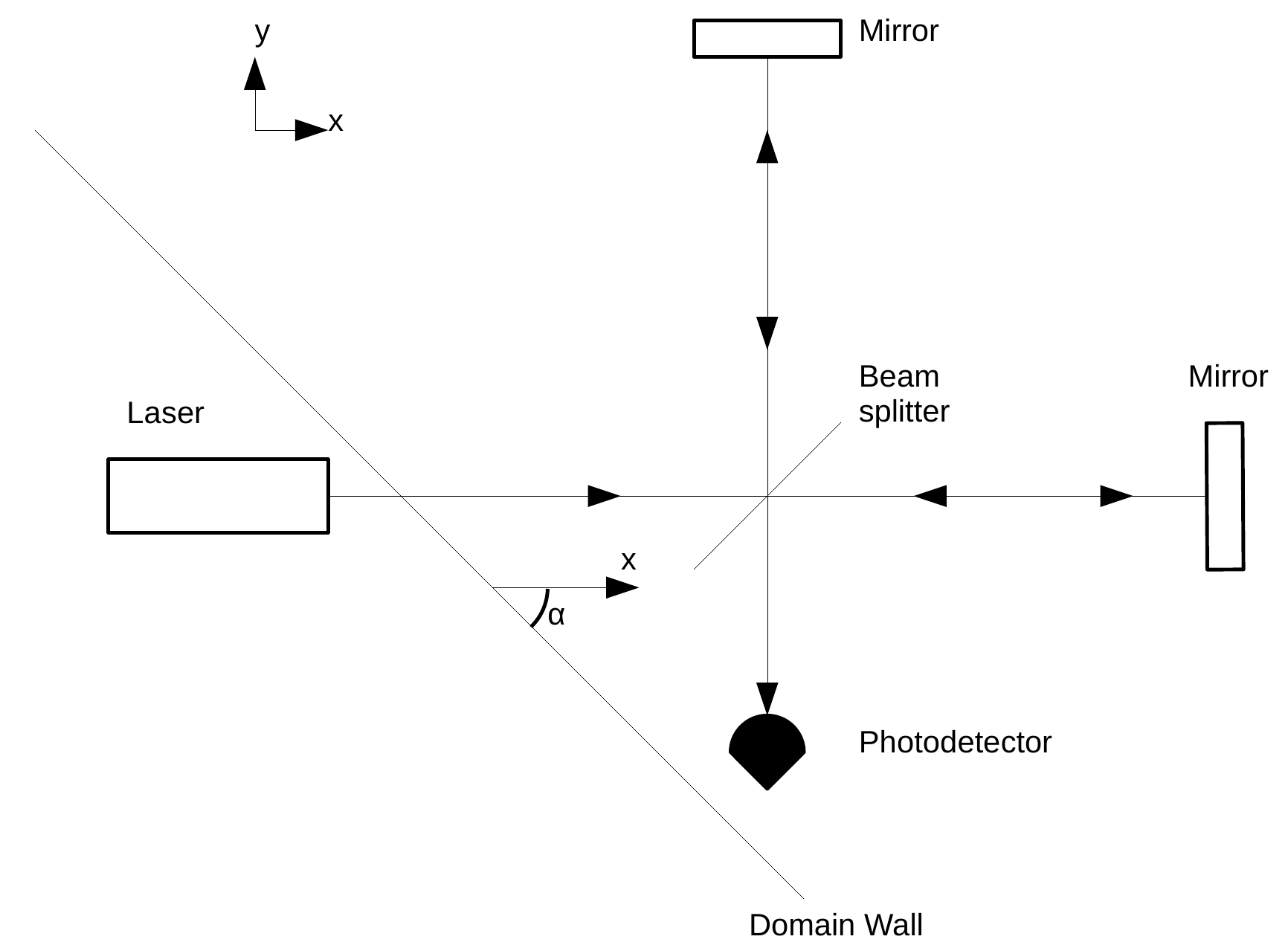}
\caption{A simplified representation of a Michelson interferometer used in gravitational wave detectors such as LIGO and VIRGO. Note that ideal mirrors produce an overall phase shift of $\pi$ which will cancel between the two arms. We also show here the detector geometry and the definition of the angle $\alpha$. We also point out here that the domain wall has a finite width $\sim2/m\sim$ few km for the masses considered in this paper. This means that the domain wall may cause additional effects in the detector apparatus (this is addressed in Section~\ref{Sec:FFSignal}).}
\label{fig:Decgeom}
\end{figure}

To briefly explain the detector geometry, a typical Michelson interferometer, as shown in Fig.~\ref{fig:Decgeom}, consists of two arms perpendicular to each other, an effect then induces a phase difference between light in the two arms causing a variation in the power output at the point where the arms join due to the superposition of the two beams.  Note, that an ideal mirror produces an overall phase shift of $\pi$ which will cancel between the two arms. In this section we take the angle $\alpha$ to be the angle the domain wall makes with the x-axis (and we assume the normal is perpendicular to the z-axis) and we align our and x- and y-axes with the detector arms. The domain wall profile is given by 
\begin{equation}
\label{eq:DWdist}
\phi \,=\, \phi_{\rm cl}({\bf x}\cdot {\bf n} -z_0 - vt)\,,
\end{equation}
where ${\bf n} = (\sin \alpha, \cos \alpha)$ is the unit normal vector to the wall, $z_0$ is the distance from the wall to the origin at $t=0$, $v$ is the velocity of the domain wall and $\phi_{\rm cl}$ is the domain wall solution profile \eqref{eq:DWsolution}.

\medskip

We can now use the dispersion relation \eqref{eq:massdisrel} to find
the phase velocity $v_{P}={\omega}/{k}$ of the photon. For a light domain wall, $m\ll k$, we have,
\begin{equation}
\label{eq:vp1}
v_{P}=\frac{\omega}{k}=1+\frac{m_{0}^{2}}{2k^{2}}\sin^{2}\left(\frac{N_{A}\phi}{f}\right).
\end{equation}
Recalling that $v_{P}=\frac{dx}{dt}$, we can then integrate both sides of this equation to get
\begin{equation}
\label{eq:2L1st}
2L=t-t_{0}+\frac{m_{0}^{2}}{2\omega^{2}}\int_{t_{0}}^{t}\,dt\,\sin^{2}\left(\frac{N_{A}\phi}{f}\right),
\end{equation}
where $t_0$ is the initial time when the light emitted by the laser arrives at the beam splitter at the beginning of the detector arm of length $L$, and $t$ is the time when the light returns to the beam splitter having travelled along the detector arm to the mirror and back (thus covering the overall distance $2L$). Note that we have replaced $k$ by  $\omega$ in the second term as to zeroth order in $m_{0}$, $\omega=k$. Now also note that to zeroth order in $m_{0}$, $t_{0}=t-2L$ so if we only wish to account for $\mathcal{O}\left(m_{0}^{2}\right)$ effects then in the limits of the integral we can replace $t_{0}$ by $t-2L$ since this term is already proportional to $m_{0}^{2}$. Hence we have,
\begin{gather}
\label{eq:2L2nd}
2L=t-t_{0}+\frac{m_{0}^{2}}{2\omega^{2}}\int_{t-2L}^{t-L}\,dt'\,\sin^{2}\left(\frac{N_{A}\,\phi_{\rm cl}((t'-t+2L)\sin\alpha-vt')}{f}\right)+ \\
\frac{m_{0}^{2}}{2\omega^{2}}\int_{t-L}^{t}\,dt'\,\sin^{2}\left(\frac{N_{A}\,\phi_{\rm cl}((t-t')\sin\alpha-vt')}{f}\right),
\end{gather}
where we have  accounted for the fact that the photon is travelling along the $X$-arm of the detector, hence in the domain wall profile \eqref{eq:DWdist}
we take $x$ to be the distance of a photon from the beam splitter at time $t$, we take $y=0$, and we placed the wall at the origin at $t=0$, hence $z_0=0$. Also note the necessity of splitting the domain of the integral into the region where the light is travelling away from/towards the detector.

For our purposes it is more useful to say that if a beam of light is detected at time $t$ then it must have been emitted at a time:
\begin{gather}
\label{eq:timeEmit}
t_{0}=t-2L+\frac{m_{0}^{2}}{2\omega^{2}}\int_{t-2L}^{t-L}\,dt'\,\sin^{2}\left(\frac{N_{A}\,\phi_{\rm cl}((t'-t+2L)\sin\alpha-vt')}{f}\right)+ \\
\frac{m_{0}^{2}}{2\omega^{2}}\int_{t-L}^{t}\,dt'\,\sin^{2}\left(\frac{N_{A}\,\phi_{\rm cl}((t-t')\sin\alpha-vt')}{f}\right).
\end{gather}

\medskip

Importantly, gravitational wave interferometers  make use of Fabry-Perot cavities and as a result light makes more than one round trip down the detector arms. 
The mean number of round trips in the cavity is related to the finesse, where, for LIGO, we have $\mathcal{F}=450$ \cite{LIGOScientific:2014pky}. We may also work with the mean amount of time light spends in the cavity, known as the storage time, $\tau_{s}=\frac{L\mathcal{F}}{\pi c}$. 
Assuming the plane-wave representation for the photon electric field, $\mathbf{E}\propto e^{-i\omega t}$,
we see that the phase difference induced by the domain wall for the light travelling along the $X$-arm of the detector is,\footnote{Note that more accurately the observed light is a superposition of beams of light which have made different numbers of round trips but we neglect this effect here.}
\begin{equation}
\label{eq:massPhaseX}
\Delta\varphi_x (t)\,=\, \frac{m_{0}^{2}}{2\omega}\int_{t-\tau_{s}}^{t}\,dt'\,\sin^{2}\left(\frac{N_{A}\, \phi_{\rm cl}  (x(t') \sin \alpha -vt')}{f}\right),
\end{equation}
and for the $Y$-arm it is,
\begin{equation}
\label{eq:massPhaseY}
\Delta\varphi_y (t)\,=\, \frac{m_{0}^{2}}{2\omega}\int_{t-\tau_{s}}^{t}\,dt'\,\sin^{2}\left(\frac{N_{A}\, \phi_{\rm cl} (y(t') \cos \alpha -vt')}{f}\right),
\end{equation}
with the domain wall solution  given by \eqref{eq:DWsolution}. Note that by $x(t)$ and $y(t)$ we mean the position of the light in the respective cavity, which will involve an appropriate splitting of the domain of the integral into regions where the light is moving away from/towards the detector as in Eq.~\ref{eq:2L2nd}.
The observable at LIGO is then the photon phase difference between the X- and Y-arms of the detector, 
\begin{equation}
\Delta\varphi = \Delta\varphi_{x}-\Delta\varphi_{y}\,.
\end{equation}
 The difference between the two arms arises due to different distances to the domain wall.

\begin{figure}
\includegraphics[width=\textwidth,keepaspectratio]{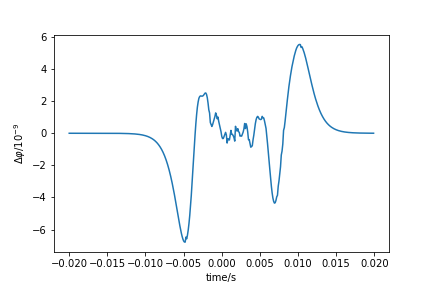}
\caption{The photon phase difference, $\Delta\varphi$, between the X- and Y-arms of gravitational waves detectors induced by the domain wall-photon mass term interaction~\eqref{eq:Lagrangian}. The phase difference is plotted as a function of time with the domain wall assumed to be at the origin at $t=0$. The chosen parameters are $m_{0}= 0.1 \,\textrm{neV},\, N_{A}=20,\, N_{\phi}=4,\, v=3\times10^{-3},\, \omega=1 \,\textrm{eV},\, L= 4 \,\textrm{km},\,\alpha=\pi/2.2,\,\mathcal{F}=450$ and $m=0.1$ neV. Note that the result is independent of the value of $f$.}
\label{fig:massresult}
\end{figure}

Finally, before presenting results we briefly discuss the value of various parameters. We take $v\sim10^{-3}$, as a generic galactic velocity, in all scenarios considered in this paper $f$ cancels and therefore the signal is independent of $f$. However it should be noted that  that $f$ should be less than $\sim1$ TeV to avoid some astrophysical constraints and, similarly, we require $m_{0}\lesssim$ few neV for astrophysical reasons as discussed in \cite{Jaeckel:2016jlh}. We plot the results in Fig. \ref{fig:massresult}.

We also wish to examine in which areas of parameter space LIGO is sensitive to a domain wall signal. In \cite{Jaeckel:2016jlh}, it was shown that LIGO is sensitive to phase shifts of order $\sim10^{-10}$. Again, as in \cite{Jaeckel:2016jlh}, we can make the transformation $\hat{x}=mx$ and $\hat{y}=my$ in Eqs.~\eqref{eq:massPhaseX}-\eqref{eq:massPhaseY} so that our integration variables are dimensionless. We then see that the phase shift is proportional to $\frac{m_{0}^{2}}{m\omega}$. So in order for a particular domain wall model to produce an observable signal at LIGO we would require $\frac{m_{0}^{2}}{m\omega}\gtrsim10^{-10}$.

\section{\label{Sec:FFDualSignal}Signal from an axion-like coupling}

In this section we consider an axion-like interaction of the domain wall with the photon given by the $\phi$ coupling to the dimension-4 photon operator via,
\begin{equation}
\label{eq:pseudoLagrangian}
\mathcal{L}_{\textrm{int}}=\frac{\tilde{g}_{DW}}{4f}\, \phi \, F_{\mu\nu}\tilde{F}^{\mu\nu}\,,
\end{equation}
where $\tilde{F}^{\mu\nu}=\frac{1}{2}\epsilon^{\mu\nu\rho\sigma}F_{\rho\sigma}$ is the dual field-strength tensor. 
The Euler-Lagrange equations for the QED photon field modified by the interaction Lagrangian \eqref{eq:pseudoLagrangian} are given by
\begin{equation}
\partial_{\mu}\left(F^{\mu\nu}+\frac{\tilde{g}_{DW}\phi}{f}\tilde{F}^{\mu\nu}\right)=0.
\end{equation}
This is equivalent to modified Maxwell's equations of the form,
\begin{gather}
\nabla\cdot\bold{B}=0 \,,\\
\nabla\times\bold{E}+\dot{\bold{B}}=0\,,\\
\nabla\cdot\bold{E}=\frac{\tilde{g}_{DW}}{f}\nabla\phi\cdot\bold{B}\,, \\
\nabla\times\bold{B}-\dot{\bold{E}}=\frac{\tilde{g}_{DW}}{f}\left(\bold{E}\times\nabla\phi-\bold{B}\dot{\phi}\right).
\end{gather}
It was shown in \cite{Harari:1992ea} that this leads to the rotation of the plane of polarisation of linearly polarised light. However for our purposes it is more useful to instead find the different dispersion relations for left and right circular polarisations which cause this rotation as was done in a similar search for axions \cite{DeRocco:2018jwe}. We will consider an experiment of the form proposed in \cite{DeRocco:2018jwe} in this section where one arm of the detector consists of right circularly polarised light and the other arm left circularly-polarised light. 

With these new Maxwell's equations the wave equations for the electric and magnetic fields become \cite{Blas:2019qqp},
\begin{gather}
\Box E_{i}+\frac{\tilde{g}_{DW}}{f}\left(\partial_{t}\phi\,\partial_{t}B_{i}-\partial_{j}\phi\,\partial_{j}B_{i}\right)=0\,, \label{eq:Eax} \\
\Box B_{i}+\frac{\tilde{g}_{DW}}{f}\left(\partial_{t}\phi\,\partial_{t}E_{i}-\partial_{j}\phi\,\partial_{j}E_{i}\right)=0\,. \label{eq:Bax}
\end{gather}
Note that the derivation of these equations requires the assumption that second derivatives of the domain wall field are negligible compared to first derivatives, and similarly first derivatives squared are negligible. We shall examine this assumption at the end of this section. Substituting the usual plane-wave ansatz for the electric and magnetic fields $\propto e^{-i\omega t+i{\bf k}.{\bf x}}$ into these equations leads to a set of equations of the form,
\begin{equation}
M(\bold{k},\omega)(\bold{E},\bold{B})^{T}=0
\end{equation}
for a particular matrix $M$. This equation has non-trivial solutions only when det~$M$=0. Imposing this condition yields the relation \cite{Blas:2019qqp}:
\begin{equation}
\omega^{2}-k^{2}=\pm \frac{\tilde{g}_{DW}}{f}(\omega\partial_{t}\phi+\mathbf{k}.\nabla\phi).
\end{equation}
These are our two solutions for left and right circular polarised light, giving the phase velocity (c.f.~\eqref{eq:vp1}),
\begin{equation}
\label{eq:vp2}
v_{P}\,=\,1\pm \frac{\tilde{g}_{DW}}{2fk^{2}}(\omega\, \partial_{t}\phi+\mathbf{k}.\nabla\phi)\,.
\end{equation}

Now we have $v_{P}=\frac{dx}{dt}$ so for a single round trip integration around the detector arm we have,
\begin{equation}
\label{eq:pseudophasex}
\int_0^{2L} dx \,=\, \int_{t_0}^t dt\,\left(1\pm \frac{\tilde{g}_{DW}}{2fk^{2}}(\omega\,\partial_{t}\phi+\mathbf{k}.\nabla\phi)\right).
\end{equation}
Noting that the second term of the right hand side is a total derivative ($\frac{d\phi}{dt}=\partial_{t}\phi+\nabla\phi.\hat{\bf{n}}$ where $\hat{\bf{n}}$ is a unit vector in the direction of propogation, $\bold{k}/k$ in this case) and performing the integration we obtain,
\begin{equation}
\label{eq:pseudophasey}
2L\,=\, t-t_{0}\pm\frac{\tilde{g}_{DW}}{2fk}\left(\phi_{\rm cl} (vt) - \phi_{\rm cl} (vt_0)\right),
\end{equation}
where $L$ is the length of a detector arm, $t$ is the time light is detected and $t_{0}$ is the time of emission. Using a similar argument as that leading to Eq. \eqref{eq:massPhaseX} we obtain:
\begin{equation}
\label{eq:pseudoPhase}
\Delta\varphi_{x/y} \,=\, \pm\frac{\tilde{g}_{DW}}{2f} \left(\phi_{\rm cl} (vt) - \phi_{\rm cl} (vt-v\tau_s)\right),
\end{equation}
with the $\pm$ applying to the different arms of the detector and the observable $\Delta\varphi$ being the difference between the two. Note that here it is the different polarisations of light that lead to the phase difference and not the different distances to the domain wall since the observable depends only on the field strength at the observation/emission points. It is worth bearing in mind here that since different polarisations of light do not interfere, we must use a waveplate to change our circularly polarised light to linearly polarised light before interfering the beams. There are also some other waveplates required in this new setup to account for the effects of reflection upon polarisation, for details see \cite{DeRocco:2018jwe}.

\medskip

\begin{figure}
\includegraphics[width=\columnwidth]{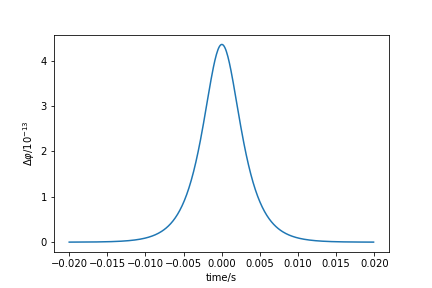}
\caption{The value of the photon phase difference $\Delta\varphi$ to be observed in the interferometer for the case of the axion-like interaction~\eqref{eq:pseudoLagrangian}
between the domain wall and photons using an experimental setup similar to that suggested in \cite{DeRocco:2018jwe}.
$\Delta\varphi$ is
plotted as a function of time for $v=3\times10^{-3},\, N_{\phi}=4,\,L=0.04\, \textrm{km},\,\tilde{g}_{DW}=10^{-10},\, \mathcal{F}=450$ and 
$m=0.1$ neV. Note that the result is independent of the value of $f$.}
\label{fig:pseudoResult}
\end{figure}

\begin{figure}
\includegraphics[width=\columnwidth]{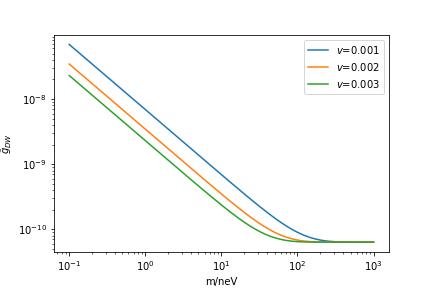}
\caption{A plot showing the sensitivity of the proposed interferometer ($L=40$ m) to domain walls for various values of the domain wall velocity, $v$. Note that the plot levels off at high masses; this is when the domain wall becomes thin enough such that a photon experiences the full passage of the domain wall in its storage time and so going to higher masses does not increase the phase shift past this point. Note that a interferometer with longer arms would improve the sensitivity to low mass domain walls. The precise value of the CAST constraint depends on the value of $f$ (whereas our phase shift doesn't) but one should bear in mind that we also require $\frac{\tilde{g}_{DW}}{f}\lesssim10^{-10}$ $\textrm{GeV}^{-1}$.}
\label{fig:axionConstraint}
\end{figure}

Let us also consider any constraints on $\tilde{g}_{DW}$. One might expect that many of the constraints on a similar axion coupling would also be applicable here, however the most stringent axion constraints come from the CAST experiment \cite{Zioutas:2004hi} which examines the solar axion flux. Although we would not expect domain walls to be produced in the sun, we would expect excitations of the $\phi$ field to be produced. These excitations would then have the possibility to convert into photons, the effect which is searched for by the CAST experiment. Therefore in presenting results we will take into account CAST constraints of $\tilde{g}_{DW}/f\lesssim10^{-10}\,\rm{GeV}^{-1}$. Results calculated using Eq.~\eqref{eq:pseudoPhase} are presented in Fig.~\ref{fig:pseudoResult}. It is worth pointing out here that these results are independent of the direction of domain wall propagation since Eq.~\eqref{eq:pseudoPhase} only contains the value of the domain wall field at the detector. In Fig.~\ref{fig:axionConstraint} we show a plot examining the values of $m$ and $\tilde{g}_{DW}$ to which we are sensitive.

\medskip

\begin{figure}
\includegraphics[width=\columnwidth]{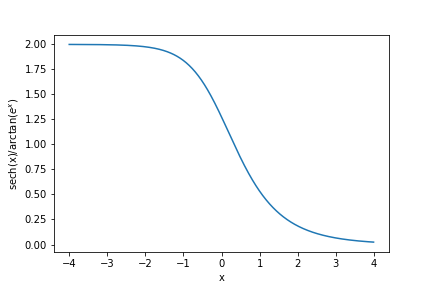}
\caption{A plot of $\textrm{sech}(x)/\arctan(e^{x})$.}
\label{fig:derivative}
\end{figure}

Finally let us examine our requirement that higher order derivatives (and higher powers of first derivatives) are negligible. This is equivalent to the requirement that \cite{Blas:2019qqp}
\begin{equation}
\label{eq:assumption}
\bigg\lvert\frac{\partial_{\mu}\phi}{\phi}\bigg\rvert\ll\omega.
\end{equation}
Taking the domain wall field in Eq.~\eqref{eq:DWsolution} and assuming propagation parallel to the z-axis (a change in direction will just introduce factors of cos/sin$\,\alpha$, where $\alpha$ is the angle between the direction of propagation and the z-axis) we obtain
\begin{gather}
\frac{\partial_{z}\phi_{\rm cl}}{\phi_{\rm cl}}=\frac{m\,\textrm{sech}(m(z-vt))}{\arctan\left(e^{m(z-vt)}\right)} \\
\frac{\partial_{t}\phi_{\rm cl}}{\phi_{\rm cl}}=\frac{mv\,\textrm{sech}(m(z-vt))}{\arctan\left(e^{m(z-vt)}\right)}. 
\end{gather}
In Fig~\ref{fig:derivative}, we plot $\textrm{sech}(x)/\textrm{arctan}(e^{x})$ to show that it is always less than two. Hence the requirement that $\big\lvert\frac{\partial_{\mu}\phi}{\phi}\rvert\ll\omega$ is equivalent to requiring that $2m\ll\omega$. This is easily satisfied for the range of $m$ and $\omega$ considered in this paper. Note that the derivative with respect to $t$ simply introduces an extra factor of $v$ and, since $v=\mathcal{O}(10^{-3})$, this does not affect the argument.

\section{\label{Sec:FFSignal}Signal from a canonical kinetic term coupling}

Now we consider the signal from an interaction of the domain wall with the photon kinetic term,
\begin{equation}
\label{eq:scalarlagrangian}
\mathcal{L}_{\textrm{int}}=\frac{g_{DW}}{4f^{2}}\phi^{2} F_{\mu\nu}F^{\mu\nu}\,.
\end{equation}
Note that since the domain wall field is a pseudoscalar, the coupling to the canonical kinetic term can involve only even powers of $\phi/f$.
Here for simplicity we concentrate on the $\phi^{2}$ interactions, but our analysis below can be easily applied to the general case of any even function of $\phi/f$.  We also note that the contribution from $\phi^{2}$ is not necessarily negligible in the EFT expansion since we expect $\phi/f=\mathcal{O}(1)$ (c.f. Eq.~\eqref{eq:DWsolution}). The Euler-Lagrange equations for the photon now read:
\begin{equation}
\partial_{\mu}\left(\left(-1+\frac{g_{DW}}{f^{2}}\phi^{2}\right){F}^{\mu\nu}\right)=0.
\end{equation}
Such interactions were considered for a fundamental particle in a strong magnetic field in e.g.\cite{Maiani:1986md,Liao:2007nu} but here we attempt to derive more general results which may be applied to interferometers where there is no magnetic field present. Instead we start, as in the previous section, with the modified Maxwell's equations resulting from the interaction in \eqref{eq:scalarlagrangian},
\begin{equation}
\nabla.\bf{B}=0\,,
\end{equation}
\begin{equation}
\nabla\times\bf{E}+\dot{\bf{B}}=0\,,
\end{equation}
\begin{equation}
\nabla.\left(\left(1-\frac{g_{DW}\phi^{2}}{f^{2}}\right)\bf{E}\right)=0\,,
\end{equation}
\begin{equation}
\nabla\times\left(\left(1-\frac{g_{DW}\phi^{2}}{f^{2}}\right)\bf{B}\right)=\frac{\partial}{\partial t}\left(\left(1-\frac{g_{DW}\phi^{2}}{f^{2}}\right)\bf{E}\right)\,.
\end{equation}
As in the previous section we neglect second derivatives and the square of first derivatives. We can then obtain the wave equations for the electric and magnetic fields (c.f. our earlier equations~\eqref{eq:Eax}-\eqref{eq:Bax} for axion-like interactions):
\begin{gather}
\Box E_{i}-\frac{g_{DW}}{f^{2}}\left(\partial_{t}\phi^{2}\partial_{t}E_{i}-\partial_{j}\phi^{2}\partial_{j}E_{i}\right)=0\,, \\
\Box B_{i}-\frac{g_{DW}}{f^{2}}\left(\partial_{t}\phi^{2}\partial_{t}B_{i}-\partial_{j}\phi^{2}\partial_{j}B_{i}\right)=0\,,
\end{gather}
where $j$ runs over spatial indices and there is an implied summation over $j$. 

\begin{figure*}
\includegraphics[width=0.5\textwidth]{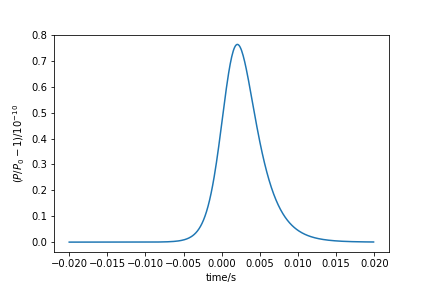}
\includegraphics[width=0.5\textwidth]{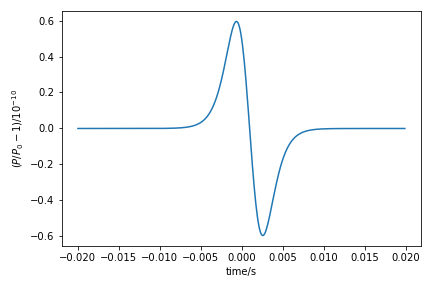}
\caption{The value of $\frac{P}{P_{0}}-1$ plotted for $v=3\times10^{-3},\, \,N_{\phi}=4,\, L= 4\, \textrm{km},\, \mathcal{F}=450,\,g_{DW}=10^{-10}$ and $m=0.1$ neV in the case where the domain wall is coupled to the photon kinetic  term via (left) \eqref{eq:scalarlagrangian}, (right) \eqref{eq:sinCoupling}. On the right we additionally have $N_{A}=2$. Here $P$ represents the observed power and $P_{0}$ is the power which would be observed in the absence of a domain wall (for a single beam). Note that the result is independent of the value of $f$.}
\label{fig:scalarresults}
\end{figure*}

Substituting in a plane-wave ansatz for the $E$ and $B$ fields,
\begin{equation}
\mathbf{E} = \mathbf{E_{0}}\,   e^{-i\omega t+i{\bf k}.{\mathbf r}}\,, \quad
\mathbf{B} = \mathbf{B_{0}}\,   e^{-i\omega t+i{\bf k}.{\mathbf r}}\,,
\end{equation}
gives the dispersion relation,
\begin{equation}
\label{eq:disrel}
\omega^{2}-k^{2}=i\frac{g_{DW}}{f^{2}}\left(\omega\partial_{t}\phi^{2}+\mathbf{k}.\nabla\phi^{2}\right).
\end{equation}
Solving for $\omega$ yields:
\begin{equation}
\omega=\lvert k\rvert+i\frac{g_{DW}}{2k f^{2}}\left(\omega\partial_{0}\phi^{2}+\mathbf{k}.\nabla\phi^{2}\right).
\end{equation}
Substituting this into the usual plane-wave ansatz we obtain
\begin{equation}
\mathbf{E}\,=\, \mathbf{E_{0}}\, e^{\,\frac{g_{DW}}{2k f^{2}}\left(\omega\partial_{0}\phi^{2}+\mathbf{k}.\nabla\phi^{2}\right)\,t}e^{-i\lvert k\rvert t+i\bold{k}\cdot \bold{r}}.
\end{equation}
It can clearly be seen that this is an ordinary plane-wave with an exponentially growing amplitude. To aid in our treatment of cavities we employ the same approach as in the previous sections and calculate the phase velocity as:
\begin{equation}
v_{P}=1+i\frac{g_{DW}}{2k^{2}f^2}\left(\omega\partial_{0}\phi^{2}+\mathbf{k}.\nabla\phi^{2}\right).
\end{equation}
Integrating as in the previous section gives
\begin{equation}
2L=t-t_{0}+ i\frac{g_{DW}}{2k f^{2}}\, \left(\phi_{\rm cl}^{2}(vt) -\phi_{\rm cl}^{2}(vt_0)\right),
\end{equation}
or
\begin{equation}
t_{0}=t-2L+i\frac{g_{DW}}{2k f^{2}}\, \left(\phi_{\rm cl}^{2}(vt) -\phi_{\rm cl}^{2}(vt_0)\right).
\end{equation}
Note that here we cannot speak of a phase difference anymore but substituting this solution into a plane-wave ansatz ($\mathbf{E}\propto e^{-i\omega t}$) gives:
\begin{equation}
\mathbf{E}\, =\, \mathbf{E_{0}}\,e^{\,\frac{g_{DW}}{2f^{2}}\phi_{\rm cl}^{2}\rvert_{t-\tau_{s}}^{t}}
\end{equation}
or equivalently, the power in the detector,
\begin{equation}
\label{eq:scalarPower}
P\,=\, |E_x^2+E_y^2| \,\propto\, e^{\,\frac{g_{DW}}{f^{2}}\phi_{\rm cl}^{2}\rvert_{t-\tau_{s}}^{t}}.
\end{equation}
In the presence of a domain wall, the interaction term in Eq.~\eqref{eq:scalarlagrangian} gives an exponentially growing power (although as we will discuss shortly, $g_{DW}$ must be small so the exponential enhancement will not be all that large). In order to examine this, in Fig.~\ref{fig:scalarresults} we plot $P/P_{0}-1$ where $P_{0}$ is the observed power in the absence of domain walls (we subtract one in order to better see the small deviations in the power). Also note that this extra energy is supplied/absorbed (depending on the sign of $\Delta\phi^{2}$) by the kinetic energy of the domain wall.

In an interferometer such as those considered in the previous sections, the light in the two different arms will be affected equally (and the two different polarisations no longer obey different dispersion relations). This will produce a fluctuation in the power output at the detector but it will in general not be possible to distinguish this signal from one such as in Section~\ref{Sec:massterm}. However a coupling of this form will uniquely produce a power change in a single beam of light since the change in power output is an intrinsic property of the domain wall and not a consequence of a phase change. This would be one possible method of determining the form of the coupling.


Before presenting results let us first examine some constraints on the value of $g_{DW}$. Adding an interaction term of the form Eq.~\eqref{eq:scalarlagrangian} leads to a change in the QED coupling constant:
\begin{equation}
\alpha\rightarrow\frac{\alpha}{1-g_{DW}\, \frac{\phi^{2}}{f^{2}}}.
\end{equation}
Since we can expect $\phi/f$ to be $\mathcal{O}(1)$, this can result in very large changes to the value of the fine structure constant. Furthermore, since a domain wall is an "event" rather than a continuously varying field, this would lead to the value of the fine structure constant on Earth changing value every time a domain wall passed. Clearly this would be catastrophic; to avoid this issue we choose $g_{DW}$ so that the change in $\alpha$ falls within the current uncertainties of its measurement. Since $\alpha$ is currently known to a precision of approximately one part in $10^{10}$ we take $g_{DW}=10^{-10}$. As in the previous section these results are independent of the direction of domain wall travel. 

Another way to avoid the change of the fine structure constant would be to replace $\phi^{2}/f^{2}$ by some function which vanishes far from the domain wall. Taking inspiration from Section~\ref{Sec:massterm} we choose:
\begin{equation}
\label{eq:sinCoupling}
\mathcal{L}_{int}=\frac{g_{DW}}{4}\sin^{2}\left(\frac{N_{A}\phi}{f}\right)F_{\mu\nu}F^{\mu\nu}
\end{equation}
This would result in $\alpha$ changing only for a very short time as the domain wall passed. It is possible that the current methods used to measure $\alpha$ would be sensitive to these variations but we will interpret any null result as a constraint on the domain wall event rate and not on $g_{DW}$. Since the derivation of Eq.~\eqref{eq:scalarPower} made no assumptions about the form of $\phi$, to obtain a new formula, one simply replaces all occurences of $\phi^{2}/f^{2}$ by $\sin^{2}\left(\frac{N_{A}\phi}{f}\right)$. This gives
\begin{equation}
P\propto e^{\,g_{DW}\sin^{2}\left(\frac{N_{A}\phi_{\rm cl}}{f}\right)\big\rvert_{t-\tau_{s}}^{t}}.
\end{equation}
However we must also examine the assumption that second derivatives are negligible. Referring to Eq.~\ref{eq:assumption} we arrive at the requirement that 
\begin{equation}
\label{eq:sinAssumption}
4m\frac{N_{A}}{N_{\phi}}\frac{\sech\left(mx\right)}{\tan\left(4\frac{N_{A}}{N_{\phi}}\tan^{-1}\left(e^{mx}\right)\right)}\ll\omega
\end{equation}
To avoid a singularity, we must have $4\frac{N_{A}}{N_{\phi}}\tan^{-1}\left(e^{mx}\right)<\pi$. As $0<\tan^{-1}\left(e^{mx}\right)<\pi/2$ this is equivalent to requiring $\frac{N_{A}}{N_{\phi}}\leq\frac{1}{2}$. Given also our requirement that $\sin^{2}\left(\frac{N_{A}\phi_{\rm cl}}{f}\right)\rightarrow0$ as $x\rightarrow\infty$, the only value of $\frac{N_{A}}{N_{\phi}}$ which satisifes both requirements is $\frac{N_{A}}{N_{\phi}}=\frac{1}{2}$. Subsituting this value into Eq.~\ref{eq:sinAssumption} and using some trigonometric identities we obtain the requirement that
\begin{equation}
8m\tanh\left(mx\right)\ll\omega
\end{equation}
which is satisified for the range of $m$ and $\omega$ considered in this paper. Results are presented in Fig.~\ref{fig:scalarresults}.

It should be noted that a change in the fine structure constant would lead to a change in the Bohr radius and thus the length of various objects as shown in \cite{Grote:2019uvn,Stadnik:2015xbn}, which would produce a signal independently of the effect calculated above. Here we will briefly derive the effect on the power at the detector when one accounts for both effects, i.e. simultaneously the power shift coming from the intrinsic interaction with the domain wall and the phase shift caused by the change in the Bohr radius.

\begin{figure*}
\includegraphics[width=0.5\textwidth]{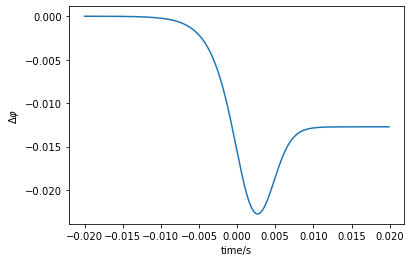}
\includegraphics[width=0.5\textwidth]{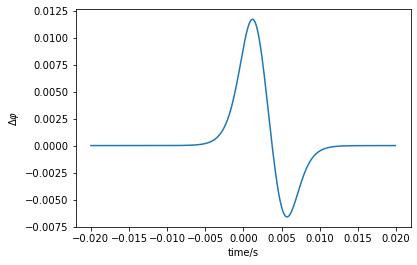}
\caption{The value of $\Delta\varphi$ plotted for $\lambda=1064\,\textrm{nm},\,\alpha=\pi/2.2,\,v=3\times10^{-3},\,n=1.45,\, \omega=1\, \textrm{eV},\,N_{\phi}=4,\, L= 4\, \textrm{km},\, \mathcal{F}=450,\,g_{DW}=10^{-10},l=6\, \textrm{cm},\,w=20\, \textrm{cm},\,m=0.1$ neV in the case where the domain wall is coupled to the photon kinetic term via (left) \eqref{eq:scalarlagrangian}, (right) \eqref{eq:sinCoupling} where we also set $N_{A}=2$. Here we plot the phase difference caused by the change in size of the beam splitter and mirrors (the power fluctuation caused by the intrinsic interaction is negligible). Note that we obtain a permanent phase shift in the left plot due to the permanent change in the value of the fine structure constant after the passage of the domain wall.}
\label{fig:scalarPhaseResults}
\end{figure*}

The change in length is given by \cite{Grote:2019uvn}:
\begin{equation}
\delta L\approx-L\frac{\delta\alpha}{\alpha}.
\end{equation}
For the range of velocities and masses considered in this paper we can take the change in length to be adiabatic. The length of the interferometer cavity is insulated from changes in length by the pendulum suspension system used in LIGO \cite{Grote:2019uvn} and we need only worry about the change in size of the beam splitter and of the mirrors which will cause a difference in optical path length. The change in optical path length is given by \cite{Grote:2019uvn}:
\begin{gather}
\delta\left(L_{x}-L_{y}\right)\approx-\sqrt{2}nl\frac{\delta\alpha}{\alpha}+\Delta w_{\textrm{x-mirror}}-\Delta w_{\textrm{y-mirror}}\approx\\
\sqrt{2}nlg_{DW}\frac{\phi^{2}}{f^{2}}\bigg\rvert_{\rm{beam-splitter}}+wg_{DW}\left(\frac{\phi^{2}}{f^{2}}\bigg\rvert_{\textrm{y-mirror}}-\frac{\phi^{2}}{f^{2}}\bigg\rvert_{\textrm{x-mirror}}\right)
\end{gather}
where $n$ is the refractive index of the beam-splitter (changes in $n$ induced by the domain wall are negligible), $l$ is the thickness of the beam-splitter and $w$ is the width of the mirrors. This must be summed over the number of round trips each photon takes, for each trip evaluating $\phi$ at the position of the beam splitter/mirrors respectively. This effect turns out to be completely dominant over the effect of the change in power. In Fig.~\ref{fig:scalarPhaseResults} we plot the phase shift caused by this effect. 

Finally for completeness we here derive the formula for the power observed at the detector in the presence of both effects, although as already noted, the effect of the term in Eq.~\ref{eq:scalarPower} is negligible compared to the phase shift induced by the change in Bohr radius. The power observed by the detector in the presence of a phase difference between the two beams is
\begin{equation}
P=\lvert A+Ae^{i\Delta\varphi}\rvert^{2}=2\lvert A\rvert^2\left(1+\cos\left(\Delta\varphi\right)\right)
\end{equation}
where $A$ is the amplitude of each beam. If we now include the effect of the intrinsic photon-domain wall interaction then the amplitude of each beam is also enhanced by a factor $e^{\,\frac{g_{DW}}{2f^{2}}\phi_{\rm cl}^{2}\rvert_{t-\tau_{s}}^{t}}$ so
\begin{gather}
P=\lvert Ae^{\,\frac{g_{DW}}{2f^{2}}\phi_{\rm cl}^{2}\rvert_{t-\tau_{s}}^{t}}+Ae^{i(\Delta\varphi_{I}+\Delta\varphi_{DW})+\frac{g_{DW}}{2f^{2}}\phi_{\rm cl}^{2}\rvert_{t-\tau_{s}}^{t}}\rvert^{2}= \\
2\lvert A\rvert^{2}e^{\,\frac{g_{DW}}{f^{2}}\phi_{\rm cl}^{2}\rvert_{t-\tau_{s}}^{t}}\left(1+\cos\left(\Delta\varphi_{I}+\Delta\varphi_{DW}\right)\right)
\end{gather}
where $\Delta\varphi_{DW}$ is the phase shift caused by the change in length of the beam splitter and mirrors and $\Delta\varphi_{I}$ is a phase shift in the absence of domain walls which is often included in the experimental design as, for various technical reasons, it makes detection easier (for details see \cite{Maggiore:2007ulw}).

\medskip

The phase shift is given approximately by
\begin{equation}
\Delta\varphi\approx\sqrt{2}nlg_{DW}\left(\frac{\mathcal{F}}{2\pi}\right)\left(\frac{2\pi}{\lambda}\right)
\end{equation}
as $\frac{\mathcal{F}}{2\pi}$ gives the mean number of round trips that light makes in the interferometer. Note that here we have neglected the effect coming from the change in width of the mirrors but as the effects are of comparable magnitude this should not have a large effect on our estimate and we have also taken $\Delta\left(\frac{\phi^{2}}{f^{2}}\right)=\Delta\sin^{2}\left(\frac{N_{A}\phi}{f}\right)=\mathcal{O}\left(1\right)$. As usual for LIGO to be sensitive to such a coupling we would require $\Delta\varphi\gtrsim10^{-10}$.

\section{Conclusion}
We briefly discuss here the prospect of detection at future, planned interferometers such as LISA. We again consider the three couplings separately. For the case of the coupling considered in Sec. \ref{Sec:massterm}, the longer interferometer arms proposed in LISA will, in general, increase the magnitude of the signal as the integrals in Eqs. \ref{eq:massPhaseX},\ref{eq:massPhaseY} will grow larger, although as with gravitational waves there is a limit to how much sensitivity can be gained purely by increasing the size of the interferometer. Similar conclusions apply to the coupling considered in Sec.~\ref{Sec:FFSignal}.

For the case of the coupling considered in Sec. \ref{Sec:FFDualSignal}, it should be noted that proposed experiments such as LISA would not be sensitive to this form of coupling since they do not use polarised light. However more generally considering the case of a larger interferometer, we point out at the end of that section that there is a maximum phase shift which can be induced by the domain wall. Therefore increasing the arm length (or equivalently increasing the storage time), is really probing lower values of $m$. 

The other main type of gravitational wave detector is a resonant bar detector. Since the domain wall model in this paper couples only to the photon it would not induce excitations of a bar as required by a resonant mass detector. However the changes in the fine structure constant could produce a signal in the case of the coupling considered in Sec.~\ref{Sec:FFSignal}. The change in length of the bar is, in principle, large enough to be detected, however the detection at a resonant bar is complex and depends on the frequency of the signal. Hence whether there is a realistic chance of detection is a more involved question requiring more detailed calculation which we leave for future work. The detection of domain walls using magnetometers was also considered in \cite{Pospelov:2012mt}, although in this paper the domain wall was coupled to fermions rather than photons.

In conclusion, we have examined the effect of domain walls on the dispersion relation of light and examined the effect of this on a signal in a Michelson interferometer. In the case of an axion-like coupling, a standard Michelson interferometer using unpolarised or linearly polarised light is insensitive to this effect and we proposed to use instead detectors designed to search for axions via a similar mechanism. The signals calculated here are within reach of current gravitational wave detectors and we have examined the parameter space for regions which current detectors woud be sensitive to.
\section*{Acknowledgements}

This work is supported by the STFC under consolidated grant ST/P001246/1 (VVK) and an STFC studentship (DLM).


\bibliographystyle{inspire}
\bibliography{main}

\clearpage
\appendix
\include{variables}

\end{document}